\documentclass[aps,pre,preprint,groupedaddress]{revtex4-2}
\usepackage{textcase}
\usepackage{graphicx,xcolor}
\usepackage{subcaption}
\usepackage{amsmath}
\usepackage{amssymb}
\usepackage[margin=3cm]{geometry}
\usepackage{natbib}
\usepackage[normalem]{ulem}

\begin{document}

\title{Non-Uniform Convergence in Moment Expansions of Integral Work Relations}


\author{Hila Katznelson}
\email{hilaweissman@campus.technion.ac.il}
\affiliation{Schulich Faculty of Chemistry, Technion-Israel Institute of Technology, Haifa 32000, Israel}

\author{Saar Rahav}
\email{rahavs@technion.ac.il}
\affiliation{Schulich Faculty of Chemistry, Technion-Israel Institute of Technology, Haifa 32000, Israel}

\begin{abstract}
	
	Exponential averages that appear in integral fluctuation theorems can be recast as a sum over moments of thermodynamic observables. We use two examples to show that such moment series can exhibit non-uniform convergence in certain singular limits. The first example is a simple model of a process with measurement and feedback. In this example, the limit of interest is that of error-free measurements. The second system we study is an ideal gas particle inside an (infinitely) fast expanding piston. Both examples show qualitative similarities; the low order moments are close to their limiting value, while high order moments strongly deviate from their limit. As the limit is approached the transition between the two groups of moments is pushed toward higher and higher moments. Our findings highlight the importance of the ordering of limits in certain non-equilibrium related calculations.	
	
\end{abstract}
\maketitle

	\section{Introduction}

The theory of stochastic thermodynamics \cite{seifert2012stochastic} offers a unifying description of out-of-equilibrium systems such as molecular motors and driven colloidal particles.
One of the main elements of the theory is the ability to consistently assign a thermodynamic observable, $X\left[ \gamma \right]$, to a single realization of a process, $\gamma$. Often, $X$ is either the work done in the process or the entropy production.
$X$ fluctuates due to random interactions with an environment and therefore the theory focuses on the distribution of $X$ in a given process, $P(X)$. Such distributions were found to satisfy several fundamental relations.

One of the major achievements of stochastic thermodynamics is a family of results known as fluctuation theorems
\cite{PhysRevLett.71.3616,PhysRevLett.74.2694,Crooks1998,Kurchan_1998,Lebowitz1999,PhysRevLett.95.040602,jarzynski1997nonequilibrium,jarzynski2011equalities}.
Fluctuation theorems express ratios of probabilities of (possibly rare) conjugate events, e.g. $P(X)/P(-X)$, in terms of thermodynamic quantities.
In fact, the discovery of this family of results, which holds far from thermal equilibrium, motivated the development of stochastic thermodynamics.
The different fluctuation theorems and  work relations come with associated integral counterparts, which takes the form of exponential averages
\begin{equation}
\label{eq:expavg}
\left\langle e^{-X}\right\rangle \equiv \int d X P(x) e^{-X} = e^{-B}.
\end{equation}
$X$ is a fluctuating quantity obtained from a realization of an out-of-equilibrium process, and $B$ has a physical meaning depending on the setup of interest.
Several examples of such integral relations are known; one of them, the Jarzynski equality, describes a system that is driven away from thermal equilibrium \cite{jarzynski1997nonequilibrium}.
In that case $X$ is proportional to the work done during the process $W$, while $B$ is the difference between two equilibrium free energies $\varDelta F$ (Here and in the following we use units where $\beta=\frac{1}{k_B T}=1$ for convenience). When the process involves measurements and feedback, $X$ becomes a sum of the work and a measure of the amount of information gained from the outcome of the measurements $I$ 
\cite{PhysRevLett.104.090602,PhysRevE.82.061120}. Finally, in the steady-state fluctuation theorems $X$ is the entropy production of a realization while $B=0$ \cite{PhysRevLett.95.040602}.

In this work, we wish to highlight an interesting mathematical property of exponential averages. Specifically, we show that the associated series of moments can exhibit non-uniform convergence when some physically motivated limits are considered.
One can evaluate the exponential average in Eq. (\ref{eq:expavg}) by recasting it as a series of moments
\begin{equation}
\left\langle e^{-X} \right\rangle = \sum_{n=0}^\infty \frac{\left( -1\right)^n}{n!} M_n,
\label{eq:genmomentS}
\end{equation}
where $M_n \equiv \left\langle X^n \right\rangle$.
We assume that the process of interest depends on some physically-relevant parameter $y$. For instance, $y$ can refer to the probability to make a measurement error, the strength of coupling between two subsystems, or the rate of which a parameter is varied. 

Consider a family of processes where $y$ can approach some limiting value $y_0$, and let us denote the corresponding limiting value of the moments by $M_n (y_0)= \lim_{y \rightarrow y_0} M_{n}(y)$.
For some specific values of $y_0$ the limiting series of moments violates the integral fluctuation theorem,
\begin{equation*}
\sum_{n=0}^\infty \frac{(-1)^n}{n!} M_n (y_0) \ne e^{-B}.
\end{equation*}
This violation can happen if one is not allowed to interchange the order of the $y$-limit and summation over moments.
\begin{equation}
\lim_{y \rightarrow y_0} \sum_{n=0}^\infty \frac{(-1)^n}{n!} M_n (y)\ne \sum_{n=0}^\infty \frac{(-1)^n}{n!} M_n (y_0).
\end{equation}
Such non-commutativity of operations points to a non-uniform convergence of the moments. The moment of any given order have a well defined limit $\lim_{y \rightarrow y_0} M_n (y) = M_n (y_0) $. Yet, when examining all moments for a given value of $y$ one finds a range of moment orders $n>\tilde{n}(y)$ for which the difference $\left| M_n (y) -M_n (y_0)\right|$ is very large, although $y \simeq y_0$.

In this paper, we use two examples to demonstrate such non-uniform convergence appears in physically interesting processes (and limits). 
The first is a process with measurement and feedback, which was used in the past to illustrate the validity of work relations for such processes \cite{koski2014experimental,horowitz2011thermodynamic}. Here, $y$ is the probability to make a measurement error, and the
non-uniform convergence occurs in the limit of error-free measurements.
The second example is a model of a single particle of an ideal gas in an expanding piston, which was studied in detail by Lua and Grosberg \cite{lua2005practical}. In this model, the parameter of interest is the piston's velocity. Non-uniform convergence is observed in the limit of an infinitely fast moving piston. 
For both examples, we characterize the asymptotic behavior of the moments and how they affect the series on the right hand side of Eq. (\ref{eq:genmomentS}) as the parameter approaches its limiting value.

We note that the non-commutativity of limits in some exponential averages was previously pointed out in several works. Press\'{e} and Sibley \cite{presse2006ordering}, who studied a free energy perturbation theory (sudden changes of a potential), considered the limit of infinitely many particles. They found that this thermodynamic limit may not commute with the limit of infinite large changes in the potential. 
In another work, Quan and Jarzynski studied the validity of the Jarzynski relation in an expanding quantum piston \cite{Quan2012}. They identified a so-called dynamic contribution to the transition probability that is completely missed by the sudden approximation, resulting in an apparent violation of the Jarzynski relation. They then showed how a more careful calculation, for finite piston velocity, allows to recover the correct value of the exponential average. Gupta and Sabhapandit studied the partial and apparent entropy production in a specific system of two Brownian particles connected to several thermal environments, as well as to each other \cite{gu}. They found that the steady-state fluctuation theorem behaves in a non-trivial way in the limit of vanishing coupling between the particles. The results described below are built upon theirs by introducing additional examples of non-commutative limits.
Importantly, we point out that such non-commuting limits are associated with non-uniform behavior of the series of moments that is obtained by expanding the exponential average.

An important application of the Jarzynski equality is as a tool for calculating equilibrium free energies from repetitions of an out-of-equilibrium process.
However, it is now understood that such calculations can suffer from poor convergence. 
Such slow convergence originates from the bias of the exponential factor in the average, $e^{-X}$, which may assign incredibly large weights to rare realizations. A large body of work was devoted to the investigation of the convergence of calculations based on Eq. (\ref{eq:expavg}), and the development of methods to improve them \cite{Gore12564,PhysRevLett.89.180602,PhysRevE.86.041130,PhysRevLett.91.140601,PhysRevLett.100.180602,doi:10.1063/1.3486196,doi:10.1063/1.1760511,doi:10.1021/jp044556a,doi:10.1063/1.2937892,PhysRevE.96.022155}.
A physically intuitive interpretation of the underlying reasons for slow convergence was given by Jarzynski \cite{jarzynski2006rare}; the rare dominant realizations that are needed to be sampled are the typical realizations of a reversed process. Our results are tangentially related to this important application of integral fluctuation theorem, since the vicinity of
the singular limit is where convergence is most difficult.

The paper is organized as follows. In Sec. \ref{sect:vanishingerror} we study a simple example of a process with measurement and feedback. We show that non-uniform convergence occurs as the probability for measurement errors approaches zero. 
In Sec. \ref{sect:idealgas}, we study a process involving a single particle gas in an expanding piston, and analyze the asymptotic behavior of moments when the piston's velocity approaches infinity. We conclude in Section \ref{sect:discussion}.
	
	\section{An information engine in the limit of error-free measurements}

The relation between information and thermodynamics has been fascinating researchers since Maxwell's celebrated thought experiment \cite{theoryheat1871}. Further important contributions to the field were made by Szilard \cite{Szilard1929}, Landauer \cite{Landauerbook}, Bennett \cite{Bennett1982}, and others.
In an important work, Sagawa and Ueda have showed that such fundamental questions can be studied from the viewpoint of stochastic thermodynamics. Specifically, they derived an integral fluctuation theorem
for processes with measurements and feedback \cite{PhysRevLett.104.090602}. It takes the form
	\begin{equation}
	\left\langle e^{-w-I}\right\rangle=e^{-\varDelta F},
	\label{eq:JRinfo}
	\end{equation}
	with 
	\begin{equation}
	I=\ln\frac{P(x|m)}{P(x)}=\ln \frac{P(m|x)}{P(m)}.
	\label{eq: mutualinfo}
	\end{equation}
$x$ denotes the state of the system at the time of measurement, while $m$ is the outcome of the measurement. $I$ is a fluctuating measure of information. It expresses one's ability to update the knowledge on the state of the system based on the measurement outcome. 
After the measurement, feedback is applied by driving the system according to the outcome. 
Equation (\ref{eq:JRinfo}) assumes a process with a single measurement, but one can readily generalize the result for processes with many consecutive measurements \cite{PhysRevE.82.061120}. Sagawa and Ueda's work has opened a new field of research, focusing on the non-equilibrium dynamics of processes in which information is manipulated \cite{parrondo2015thermodynamics}. 

In this section we study a simple example of an out-of-equilibrium process with a measurement and feedback. This example have been used before to illustrate the validity of the integral work relation (\ref{eq:JRinfo}) \cite{koski2014experimental,horowitz2011thermodynamic}. Here our goal is to show that the moments on the right hand side of Eq. (\ref{eq:genmomentS})
converge in a non-uniform manner when the limit of error free measurements is taken.

Consider a system with two discrete states, which we term left and right ($x=L,R$, respectively). Let us assume that we can control the energies of these states ${\cal E}_{L,R} (t)$. In the process we examine, the system is initially in thermal equilibrium with temperature $T$, and the energies of the two states are ${\cal E}_L (0)={\cal E}_R (0)=0$. At time $t=0$ a measurement is made to determine if the system is in the left or right state. The measuring device is not perfect, hence occasionally it makes errors. We characterize this property using the conditional probability to find outcome $m$ when the system is in state $x$, $P(m|x)$. Specifically, we use
	\begin{eqnarray}
	P(L|L) & = & P(R|R) = 1 -\varepsilon, \\
	P(L|R) & = & P(R|L) = \varepsilon.
	\label{eq:vanisherrorprobabilities}
	\end{eqnarray}
$\varepsilon$ is the probability to make a measurement error. Note that the setup has a right-left symmetry that is used to simplify the calculations.

\begin{figure}
		\centering
		\begin{subfigure}{0.8\linewidth}
			\includegraphics[width=\linewidth]{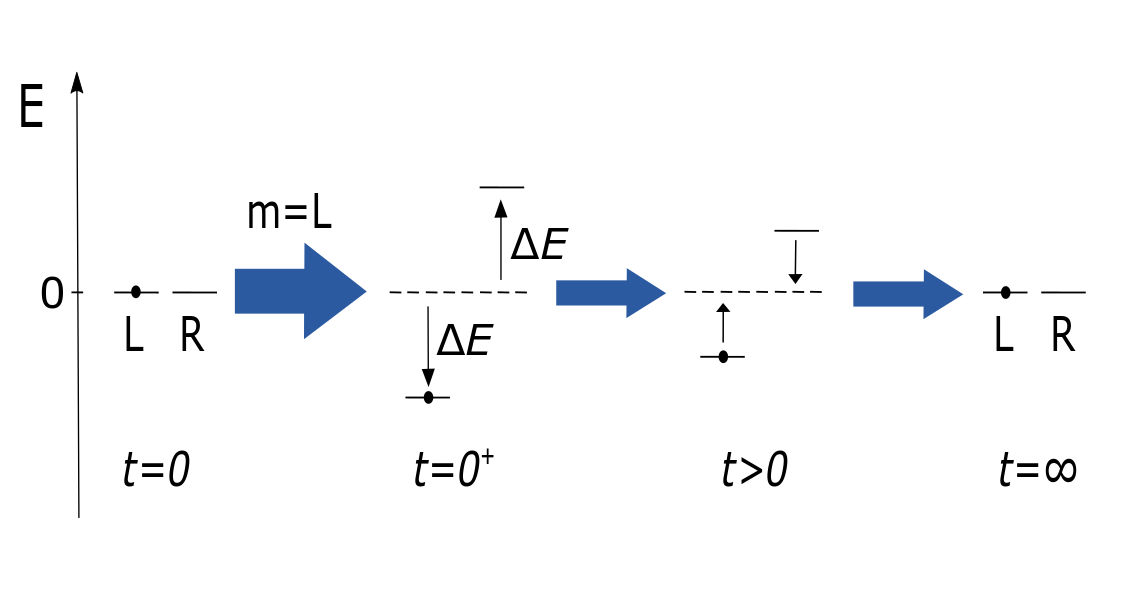}
		\end{subfigure}
		\caption{A heuristic illustration of the process. The figure depicts a realization with an accurate measurement; a measuring device measures the state of a system at time $t=0$. Feedback is applied immediately by reducing the measured state energy by $\Delta E$ and increasing the other's energy by the same among. Afterwards, the system undergoes quasistatic relaxation to its initial energies, closing the cycle.}
		\label{fig:measex}
	\end{figure}

Feedback is applied immediately following the measurement, in an attempt to extract energy from the system; specifically, the energy of the state $m$ is reduced to ${\cal E}_m (0^+)=-\Delta E$, while the energy of the other state is increased to $+\Delta E$. The energies ${\cal E}_{L,R} (t)$ are subsequently relaxed back to $0$ in a quasistatic and isothermal process to complete a cycle. For small $\varepsilon$, the probability that the system is indeed at the site $m$ at $t=0$, when the site's energy is lowered, is larger than the probability to occupy this site during the isothermal process. This aspect of the process allows to extract energy from the information gained by measurements in the cycle. A realization with an accurate measurement ($x=m=L$) is illustrated in Figure \ref{fig:measex}.
	
The work done on the system has two contributions, from the sudden change of energy, and from the quasistatic segment of the cycle. In the quasistatic segment, the work done is exactly $w=F(\infty)-F(0^+)=\ln \cosh \Delta E$.	
Here we use the fact that the work distribution of quasistatic isothermal distributions is a delta function centered at $w=\Delta F$ \cite{doi:10.1063/1.1353552,PhysRevE.70.066112}. As a result, the work done during a realization is determined by the presence or absence of a measurement error.If the measurement is accurate, the sudden change contributes $-\Delta E$ to the work. Otherwise the contribution is $+\Delta E$.
 The two possible values of work are
	\begin{equation}
	W_a=- \Delta E + \ln \cosh \Delta E,
	\label{eq:workrightmeasure}
	\end{equation}
and
	\begin{equation}
	W_i = + \Delta E + \ln \cosh \Delta E.
	\label{eq:workerrormeasure}
	\end{equation}
The value of $I_a=\ln 2 \left(1- \varepsilon \right)$ or $I_i = \ln 2 \varepsilon$, also depends on whether the measurement is accurate or not.
The need to consider only two possible values of $w+I$ makes this process easy to analyze. This is the reason it has been previously used to illustrate the validity of fluctuation relations for processes with measurement and feedback \cite{horowitz2011thermodynamic}. One can then readily verify that
	\begin{align}
	\left\langle e^{-w-I}\right\rangle& ={\sum_{x,m}P(x,m)e^{-w(x,m)-I(x,m)}=}\nonumber \\&= \left( 1-\varepsilon \right) e^{-W_a-\ln 2(1-\varepsilon)}+ \varepsilon e^{-W_i -\ln 2\varepsilon}=1,
	\label{eq:measureexpavgright}
	\end{align}
as one expects since $\varDelta F=0$ for cyclic processes.
	
We now demonstrate that this simple information engine exhibits non-uniform convergence in the limit $\varepsilon \rightarrow 0$. This non-uniformity expresses itself as sensitivity to the order in which the limit and various summations and integrations are taken. Below we use the moment expansion of the exponential average in Eq. (\ref{eq:JRinfo}) to illustrate this mathematical property.
   
Let us examine the $n^{th}$ moment of $ \langle e^{-w-I}\rangle$,
	\begin{equation}
	M_n (\varepsilon) = \left\langle[w +I ]^n\right\rangle= M_n^{(a)} +M_n^{(i)},
	\label{eq:measureerrormoment}
	\end{equation}
where $M_n^{(a)}(\varepsilon) \equiv  (1-\varepsilon) \left[W_a + \ln  2(1-\varepsilon)\right]^n$ and $M_n^{(i)}(\varepsilon)\equiv \varepsilon  \left[W_i + \ln 2\varepsilon \right]^n$.
All the moments have a well defined value in the limit  $\varepsilon \rightarrow 0$
	\begin{equation}
	M_n (0)= M_n^{(a)}(0)= (W_a+\ln 2)^n.
	\label{eq:measureerrormomentlimit}
	\end{equation}
However, substituting the limit of these moments in the series clearly leads to a wrong result since
	\begin{equation}
	\sum_{n=0}^{\infty} \frac{(-1)^n M_n (0)}{n!}=e^{-W_a-\ln 2} \ne 1.
	\label{eq:errorwrongsum}
	\end{equation}
This apparent violation of Eq. (\ref{eq:measureexpavgright}) occurs because
    \begin{equation}
    \lim_{\varepsilon\to 0}\sum_{n=0}^{\infty}\frac{(-1)^n}{n!}M_n(\varepsilon)\neq \sum_{n=0}^{\infty}\frac{(-1)^n}{n!}\lim_{\varepsilon\to 0}M_n(\varepsilon).
    \label{eq:switchordererror}
    \end{equation}

\begin{figure}
    		\centering
    		\begin{subfigure}{0.7\linewidth}
         		\includegraphics[width=\linewidth]{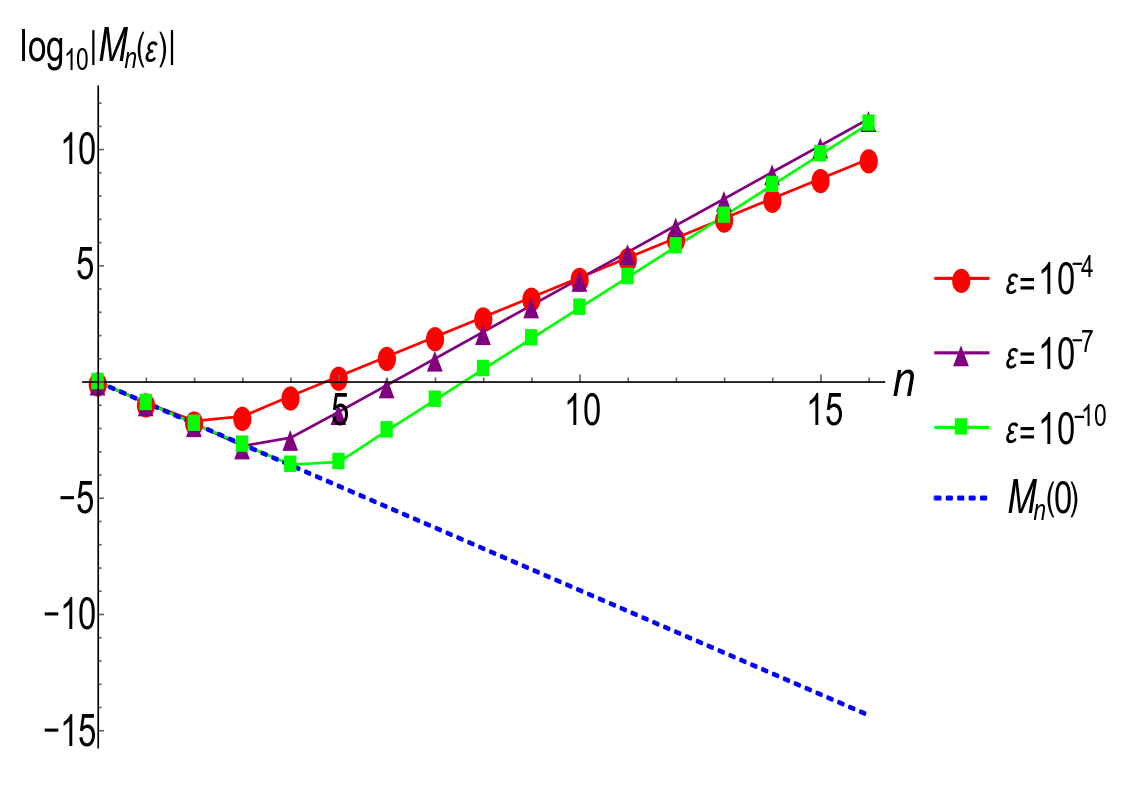}
         			\caption{}
         		    	\label{fig:logmomentsVSnVSerror}
    		\end{subfigure}
    		\begin{subfigure}{0.7\linewidth}
    	    	\includegraphics[width=\linewidth]{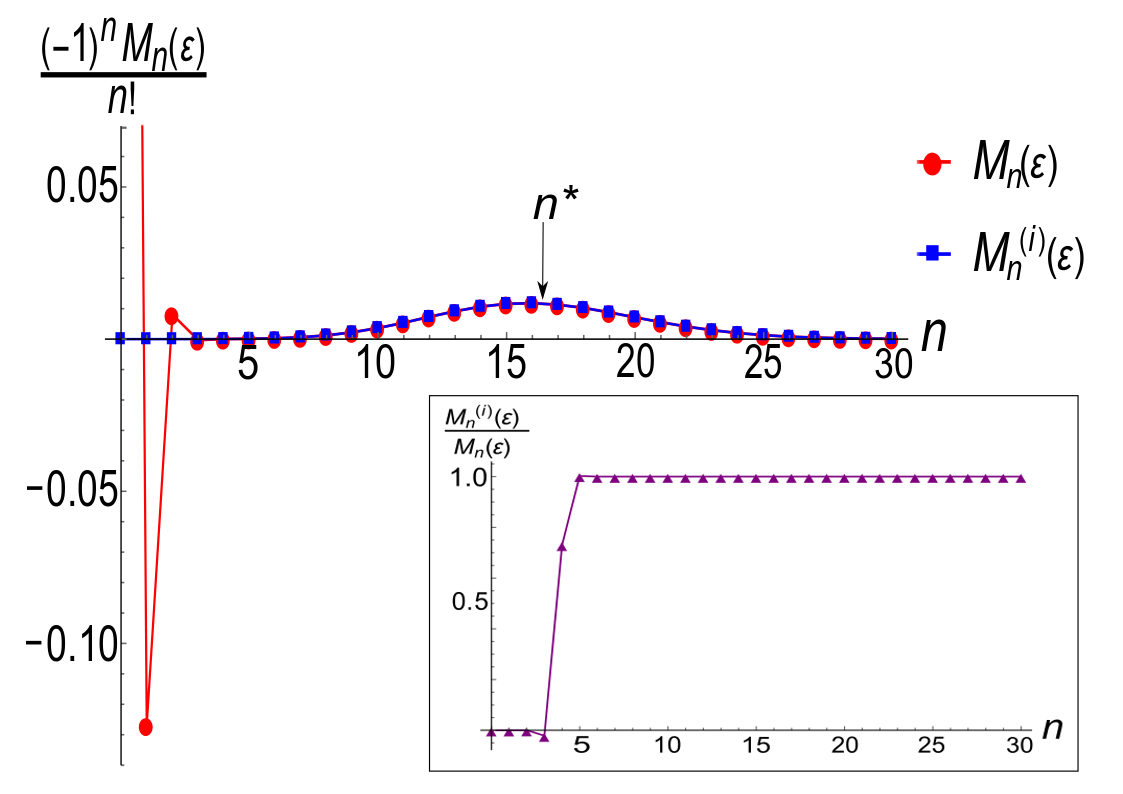}
    	        	\caption{}
    	            	\label{fig:exactVserrorterm}
    		\end{subfigure}
             \caption{(a) A logarithmic plot of the $n$'th moment, $M_n(\varepsilon)$, as a function of $n$. Different curves correspond to different values of $\varepsilon$. The dashed curve shows the limiting value of the moments, $M_n (0)$.  (b) A comparison between the terms of the series $\frac{(-1)^n}{n!} M_n (\varepsilon)$ and the corresponding contribution using only realizations with inaccurate measurements. Here $\varepsilon=10^{-8}$. $n^* \simeq 16.3$, marked by an arrow, is the simple estimate for the center of the peak (\ref{eq:peak}). Note that the $M_0(\varepsilon)$ term is out of range on this graph and is equal to 1. In both panels $\varDelta E=\beta=1$. The inset depicts the ratio $M_n^{(i)} (\varepsilon)/M_n (\varepsilon)$ as a function of $n$.}
             \label{fig:vanishingerrorresilts}
    	\end{figure}

The non-uniform convergence can be illuminated by examining the value of the 
moments as a function of both $\varepsilon$ and $n$. Figure \ref{fig:logmomentsVSnVSerror} depicts a logarithmic plot of $M_n (\varepsilon)$
as a function of $n$. Different curves correspond to different values of the measurement error $\varepsilon$. The dashed linear line is the limiting value $M_n (0)$.
$M_n (\varepsilon)$ clearly follows $M_n (0)$ for a range of $n$'s,
but starts to deviate sharply from its limiting value for large enough $n$.
Crucially, as the value of $\varepsilon$ is decreased the point at which the curves of $M_n(\varepsilon)$  start to deviate from $M_n (0)$ is pushed to higher and higher moments. 
This behavior results from terms of the form $\varepsilon \ln^n \varepsilon$ that appear in $M_n^{(i)} (\varepsilon)$. Such terms have vastly different behavior depending on which of the limits $\varepsilon \rightarrow 0$ and $n \rightarrow \infty$ is taken first. Specifically for any fixed value of $n$, $\varepsilon \ln^n \varepsilon \rightarrow 0$ when $\varepsilon \rightarrow 0$. Alternatively, $\varepsilon \left| \ln \varepsilon\right|^n \rightarrow \infty$ when $\varepsilon \ll 1$ and $n \rightarrow \infty$.

This behavior is consistent with the validity of the exponential average in Eq. (\ref{eq:JRinfo}); we demonstrate it by plotting $(-1)^n M_n (\varepsilon)/n!$ as a function of $n$.
The result is depicted in Figure \ref{fig:exactVserrorterm}. After a few points with
alternating signs the curve exhibits a fairly smooth and wide peak that is centered around high moments. The factor of $1/n!$ is important as it suppresses the contribution from even larger orders of $n$. This peak originates from the same term that causes the deviation of $M_n (\varepsilon)$ from $M_n (0)$, $\varepsilon \ln^n \varepsilon$. We infer this conclusion from the second line in the figure, $\frac{(-1)^n}{n!} M_n^{(i)}$, which is indistinguishable from the exact result in the region of the peak.This is demonstrated in the inset of Fig. \ref{fig:exactVserrorterm}, which shows the ratio of  between $M_n^{(i)} (\varepsilon)$ and $M_n (\varepsilon)$. This ratio is very close to 1 for the relevant moments.

The center of the peak can be estimated by taking the logarithm of $\frac{(-1)^n}{n!} M_n^{(i)}$ and finding its maximum. With the help of Stirling's approximation, this short calculation leads to
\begin{equation}
\label{eq:peak}
n^* \simeq -\Delta E - \ln 2 \cosh \Delta E - \ln \varepsilon.
\end{equation}
The peak's width can be estimated using a gaussian approximation
\begin{equation}
\delta n \simeq \sqrt{n^*}.
\label{eq: width}
\end{equation}
The value of $n^*$ is marked in Fig. \ref{fig:exactVserrorterm}, and is indeed fairly close to the center of the peak, as expected. Equations (\ref{eq:peak}) and (\ref{eq: width}) clarify how the peak changes as $\varepsilon$ is varied; the center of the peak moves to higher moment orders for smaller values of $\varepsilon$, and its width increases.     
     
In conclusion, the non-uniform convergence manifests itself through the behavior of the moment sequence and of Equation (\ref{eq:genmomentS}).
In particular, a parameter dependent peak is found. When the limit $\varepsilon \rightarrow 0$ is taken the center of the peak is pushed towards higher moment orders, and its width increases. The apparent violation of the exponential average results from the shift in the peak as $\varepsilon$ is decreased; taking the limit $\varepsilon \rightarrow 0$ first and only then summing over all moments misses the contribution from the peak entirely. This contribution, which ultimately comes from realizations with measurement errors, is responsible for the correct convergence of the exponential average and the validity of the work relation. 

\label{sect:vanishingerror}
\section{A particle in an expanding piston}
	
The second example we consider is that of a single gas particle in an expanding piston. This process is simple enough to be exactly solvable. It was studied by Lua and Grosberg \cite{lua2005practical}, who have calculated the work distribution of the process, and demonstrated that it satisfies the Jarzynski equality
 \begin{equation}
 \label{eq:Jequality}
 \left\langle e^{- w} \right\rangle = e^{-\Delta F}.
 \end{equation}
In this section, we use their results to discuss the behavior of moments of the work distributions in the singular limit of infinitely fast expansion.
 
\begin{figure}
     	\centering
     	\begin{subfigure}{0.6\linewidth}
     		\includegraphics[width=\linewidth]{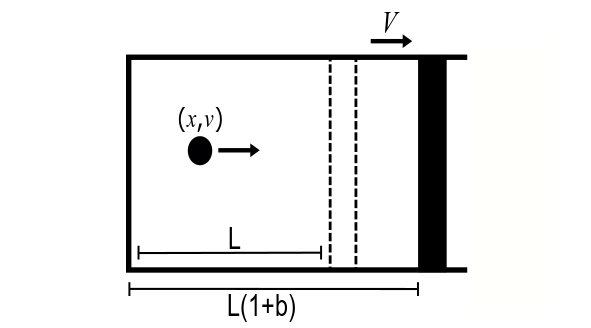}
    	\end{subfigure}
     	\caption{An illustration of the expansion process. The initial length of the cylinder is $L$, and the expansion is halted when it reaches a value of $(1+b)\cdot L$. The piston moves with a fixed velocity $V$ throughout the process. The particle is initially at point $x$, with a velocity $v$ taken out of a Maxwell-Boltzmann distribution.}
     	\label{fig:idealgasexpansion}
     \end{figure}

Consider a single gas particle in a piston of initial length $L$. The setup can be viewed effectively as one dimensional since motion in transverse directions does not play a role. Initially, the system is brought to equilibrium with a heat reservoir with inverse temperature $\beta=1$. At time $t=0$ the system is disconnected from this environment, and the piston is pushed outwards with a fixed velocity $V$. The process ends at time $t=b\cdot L/V$, when the piston reaches a size of $(1+b)L$. $b$ is an expansion factor. This process is depicted in Figure \ref{fig:idealgasexpansion}.

During the process the energy of the gas particle can change only due to collisions with the piston. The change in the kinetic energy of the gas is therefore equal to the work that is done on the piston. This work depends only on the particle's initial position $x$ and velocity $v$, and can be calculated by solving the equation of motion. It is simply expressed in terms of the initial velocity, $v$, and the number of times that the particle collides with the piston, $l$,
     \begin{equation}
     w_{l}=E_{k,f}-E_{k,i}=2l^{2}V^{2}-2l V v.
     \label{eq:idealgaswork}
     \end{equation}
For $|v|<V$  there are no collisions during the process and as a result $w=0$. In an expansion process $|v|$ decreases after each collision so that $w_l<0$. We note that our definition of work differs from Lua's and Grosberg's by a sign because they have considered the work that the particle have done on the piston as positive. 

The probability to observe a given amount of work can be calculated by integrating over all the initial conditions that lead to the same value of $l$. This problem was solved exactly by Lua and Grosberg \cite{lua2005practical}. They found that the work distribution of the expansion process is given by
     \begin{equation}
     P(w)=P_{0}\delta(w)+\sum_{l=1}^{\infty}\frac{1}{\sqrt{2\pi}lV}e^{-\frac{1}{2}(l V-\frac{w}{2lV})^{2}}f (w).
     \label{eq: wdistributionidealgas}
     \end{equation}
Each term of the summation matches a specific number of collisions $l$, while $P_{0}$ is the probability that no collisions occur in the process. The explicit expression for $P_0$ is rather cumbersome, and is not be needed for the calculations below. $f (w)$ is an overlap factor
     \begin{equation}
     f (w)=
     \begin{cases}
     -(l-1)(\frac{b}{2}+1)-\frac{bw}{4lV^{2}},  & -2lV^{2}(l-1)-\frac{4l^{2}V^{2}}{b}<w\leq -2lV^{2}(1+\frac{2}{b})(l-1)\\
     1,  &-2lV^{2}(l+1)-\frac{4l^{2}V^{2}}{b}<w\leq -2lV^{2}(l-1)-\frac{4l^{2}V^{2}}{b} \\
     (l+1)(\frac{b}{2}+1)+\frac{bw}{4lV^{2}},  & -2lV^{2}(1+\frac{2}{b})(l+1)<w<-2lV^{2}(l+1)-\frac{4l^{2}V^{2}}{b}
     \end{cases},
     \label{eq:overlap}
     \end{equation} 
that is built from separate segments originating from realizations with different numbers of collisions between the particle and piston. Note that $0\le f(w) \le 1$.

We use a slightly different parametrization of the expansion process when compared to that of Lua and Grosberg  \cite{lua2005practical}. Specifically, since we want to take the limit $V \rightarrow \infty$ while keeping the same free energy difference, the expansion process ends when the piston reaches a predetermined length of $(1+b)L$. Equations (\ref{eq: wdistributionidealgas}) and (\ref{eq:overlap}) have therefore been rewritten in a form that explicitly depends on $V$ and $b$.

Lua and Grosberg showed that this work distribution satisfies the Jarzynski equality (\ref{eq:Jequality}) \cite{lua2005practical}, which here takes the form $\left\langle e^{-w}\right\rangle=1+b$. They also noticed that the limit $V \rightarrow \infty$ is singular. 
This can be easily seen from a naive estimation of the limit. When $V \rightarrow \infty$ the probability that the particle will collide with the piston vanishes, and as a result $P(w) \rightarrow \delta \left( w \right)$. 
This naive argument suggests that
  \begin{equation*}
     \lim_{V \rightarrow \infty} \left\langle e^{-w} \right\rangle = \int d w \delta \left( w \right) e^{-w}=1.
  \end{equation*}
This incorrect result implies that one can not change the order of the $V \rightarrow \infty$ limit and integration over $w$. 
Lua and Grosberg then showed that the Jarzynski equality is restored once the far tails of $P(w)$ are taken into account. Below we show that this phenomenon can be
recast as a non-uniform convergence of the moments of the work distribution. Moreover, these moments exhibit the same qualitative behavior as the series of moments that was studied in Section \ref{sect:vanishingerror}.

We are interested in the moments of the work distribution $$M_n=\left\langle w^n \right\rangle=\int dw w^n P(w).$$
The Jarzynski equality (\ref{eq:Jequality}) can be recast as
\begin{equation}
\label{eq:momseries}
 \sum_{n=0}^\infty \frac{(-1)^n}{n!}M_n=1+b.
\end{equation}
The naive argument discussed above can be rephrased in terms of the moments. If one simply takes the $V \rightarrow \infty$ limit for each moment, one finds that $\lim_{V  \rightarrow \infty} M_n= \delta_{n0}$, so that $\lim_{V \rightarrow \infty} \sum_{n=0}^\infty \frac{(-1)^n}{n!}M_n=1$. We now turn to study the asymptotic behavior of the moments and of the series (\ref{eq:momseries}) for large $V$.

The calculation of the moments involves an integral of a product of two factors, namely $P(w)$ and $w^n$. The exponents in $P(w)$ are maximal for $w_l=2l^2 V^2 \gg 1$. These express the small probability to find collisions for large $V$, and give large weight to realizations with small $|w|$. The factor of $\left| w \right|^n$ can counteract this, by giving weight for realizations with large $|w|$. For very high moments, with $n \sim V^2$, the two terms balance each other, resulting in a dominant contribution to the integral over $w$ that comes from inside the domain of integration. It is these moments that can account for the convergence of the series (\ref{eq:momseries}).

The complicated form of $P(w)$ makes the full characterization of the asymptomatic behavior of the moments quite involved\textbf{,} since different numbers of collisions and work values may become dominant for different parameters. In our analysis we choose a regime of parameters ($b<2$) where the consideration becomes simpler. When $0<b<2$, the convergence of the series (\ref{eq:momseries}) is dominated by realizations with a single collision and work that is in the range $-\frac{4 V^2}{b} \le w \le 0$. Higher moments with $n\gg V^2$ can exhibit different asymptotic behavior, but the factor of $1/n!$ in the series (\ref{eq:momseries}) ensures that their contribution is sub-dominant.  

We therefore approximate the moments by using the component of these relevant work values in $P(w)$
\begin{equation}
\label{eq:momentappro}
 M_n \simeq - \int_{-\frac{4}{b}V^2}^0 \frac{bw^{n+1}}{4V^3\sqrt{2\pi}}e^{-\frac{1}{8}V^2(\frac{w}{V^2}-2)^2}dw. 
 \end{equation}
Eq. (\ref{eq:momentappro}) is expected to be a good approximation for the moments $n=1,2, \cdots, \tilde{n}$, where $\tilde{n} \simeq \left(2/b+4/b^2 \right) V^2$, and to break down for moments with $n>\tilde{n}$. This approximation can be improved by including contributions with additional collisions and a larger domain of work values. However, as we show below, the $n<\tilde{n}$ moments suffice for the convergence of the Jarzynski equality.
Changing variables to $y=-\frac{w}{V^2}$ results in
	\begin{equation}
	M_n \cong (-1)^n\frac{b}{4\sqrt{2\pi}}V^{2n+1} \int_{0}^{\frac{4}{b}}y^{n+1}e^{-\frac{1}{8}V^2(y+2)^2}dy.
	\label{eq:momentapproxy}
	\end{equation}
The asymptotic behavior of the integral in (\ref{eq:momentapproxy})
depends on $n$ and $V$ through the ratio $\alpha=\frac{n+1}{V2}$.
Several distinct regimes of moments are possible depending on the value of $\alpha$.
		
$\alpha \ll 1$ is obtained when one fixes the order of the moment $n$ while taking $V \rightarrow \infty$. In this case, the integral is dominated by $y$ values satisfying $y \ll 1$. This detail allows us to Taylor expand the exponent and neglect the quadratic term in $y$, giving
\begin{equation}
M_n(\alpha\ll1) \cong (-1)^n\frac{b}{\sqrt{2 \pi}} V^{2 n +1} e^{-\frac{1}{2} V^2} \int_0^{\frac{4}{b}} y^{n+1} e^{-\frac{1}{2} V^2 y} dy.
\end{equation}
Extending the upper limit of this integral to infinity leads to an integral representation of the Gamma function. One finds
	\begin{equation}
	M_n(\alpha\ll1) \cong(-1)^n\frac{b}{\sqrt{2\pi V^3}}2^n(n+1)!e^{-\frac{1}{2}V^2}.
	\label{eq:momentapm<vp}
	\end{equation}

In the second regime, $n$ is in the same order of magnitude as $V^2$, so $\alpha \sim 1$. The component $y^{n+1}$ in the integrand of Eq. (\ref{eq:momentapproxy}) becomes relevant and should not be assumed to be small in comparison to the exponent.  Thus, we rewrite Eq. (\ref{eq:momentapproxy}) as
	\begin{equation}
	M_n \cong(-1)^n\frac{b}{4\sqrt{2\pi}}V^{2n+1} \int_{0}^{\frac{4}{b}} e^{-\frac{1}{2}V^2(\frac{1}{4}(y+2)^2-2\alpha \ln y)}dy.
	\label{eq:momentappy2}
	\end{equation}
The function $\frac{1}{4}(y+2)^2-2\alpha \ln y$ in the exponent has a local minimum at $y*=-1+\sqrt{1+4\alpha}$. As long as $y^*$ is inside the domain of integration, one can obtain an approximation for the moment by following several simple steps. First, we Taylor expand this function around $y^*$ up to its second order, neglecting higher order terms, and then expand the range of integration to infinity. This standard asymptotic evaluation \cite{Benderbook} results in
	\begin{equation}
	M_n(\alpha\sim1) \cong (-1)^n\frac{b}{2}V^{2n}\sqrt{\frac{1+2\alpha-\sqrt{1+4\alpha}}{1+4\alpha-\sqrt{1+4\alpha}}}e^{-\frac{1}{2}V^2[\frac{1}{2}(1+2\alpha+\sqrt{1+4\alpha})-2\alpha ln(-1+\sqrt{1+4\alpha})]}.
	\label{eq:momentapm=v}
	\end{equation}
This approximation clearly  breaks down for $y^* \ge \frac{4}{b}$,
which is equivalent to $\alpha \ge \frac{2}{b}+\frac{4}{b^2}$.
Thus, the asymptotic expression (\ref{eq:momentapm=v}) fails for moments
with $n>V^2 \left( \frac{2}{b}+\frac{4}{b^2}\right)=\tilde{n}$.

Finally, when $\alpha \gg 1$, $y \simeq \frac{4}{b}$ becomes the dominant part of the integral as $y^*$ shifts to a value beyond the upper limit of the integral; one can naively estimate the integral in (\ref{eq:momentappy2}) by expanding around $y=\frac{4}{b}$, resulting in
	\begin{equation*}
	M_n(\alpha\gg1) \cong(-1)^n \frac{bV^{2n-1}}{2\sqrt{2\pi}(\frac{b\alpha}{2}-\frac{2}{b}-1)}e^{-\frac{1}{2}V^2[\frac{1}{4}(\frac{4}{b}+2)^2-2\alpha ln\frac{4}{b}]}.
	\end{equation*} 
However, this expression is not a good approximation of the moment $M_n$, since the neglected contributions from $w<-\frac{4}{b} V^2$ become larger then the contributions that were accounted for. Importantly, a better estimation of these moments is not needed for the main purpose of this discussion. 
As we show next, when $b<2$, the moments with $\alpha \sim 1$ are those responsible for the convergence of the Jarzynski equality.

We now examine the contributions of moments in these three regimes to the series $\sum_{n=1}^\infty \frac{(-1)^n}{n!} M_n$ when $V \gg 1$ and determine which is the most dominant.
This series must converge to $b$ because the zero'th moment already contributes 1.
First, we estimate the contribution of moments with fixed $n>0$ (the regime is $\alpha \ll 1$) and show that their contribution to the sum is small. Consider all the moments from $n=1$ to $n=\gamma V^2$, where $\gamma$ is a number 
satisfying $\gamma \ll 1$. One can then use Eq. (\ref{eq:momentapm<vp}) for these moments and obtain
	\begin{equation}
	I_{\alpha\ll1}= \sum_{n=1}^{\gamma V^2}\frac{(-1)^n}{n!}M_n \cong \sqrt{\frac{2}{\pi}}\frac{\gamma b}{V}e^{V^2(\gamma ln2-\frac{1}{2})}.
	\label{eq:momentapm<vsum}
	\end{equation}
This sum approaches $0$ as $V \rightarrow \infty$. This result just expresses the fact that every given moment in this regime satisfies $\lim_{V \rightarrow \infty} M_n =0$. It matches the convergence of Eq. (\ref{eq:Jequality}) to the wrong value of $1$ when the wrong order of limits is used.

For the regime of  $\alpha \sim 1$, the contribution of moments to the series (\ref{eq:genmomentS}) can be estimated with the help of Equation (\ref{eq:momentapm=v}).
We focus on the sum
\begin{equation}
 I_{\alpha \sim 1} \equiv \sum_{n=n_l}^{n_h} \frac{(-1)^n}{n!} M_n.
\end{equation}
Here, $n_l$ and $n_m$ are chosen so that they bracket the dominant part of the series in this sum, while also stay in range of validity of Equation (\ref{eq:momentapm=v}). We recast $I_{\alpha \sim 1}$ by using Stirling's approximation, and subsequently replace
the summation over $n$ by an integral. The latter operation is justified because discrete 
changes in $n$ are mapped to extremely small changes in $\alpha=\frac{n+1}{V^2}$.
This series of steps leads to
	\begin{equation}
	I_{\alpha\sim1}\cong \frac{b V}{2\sqrt{2\pi}}\int_{\alpha_l}^{\alpha_h}\sqrt{\alpha\frac{1+2\alpha-\sqrt{1+4\alpha}}{1+4\alpha-\sqrt{1+4\alpha}}}e^{-\frac{1}{2}V^2h(\alpha)}d\alpha,
	\label{eq:momentapm=vsum}
	\end{equation}
where
	\begin{equation}
	h(\alpha)=\frac{1}{2}(1+2\alpha+\sqrt{1+4\alpha})-2\alpha ln(-1+\sqrt{1+4\alpha})+2\alpha ln\alpha-2\alpha.
	\label{eq:hI2expfun}
	\end{equation}
The function $h(\alpha)$ has a minimum at $\alpha^*=2$ with $h(2)=0$. As a result, the integrand has a maximum there, and its value drops sharply with $|\alpha -\alpha^*|$. Thus, for any fixed values of $\alpha_l<2$ and $\alpha_h>2$, the leading order asymptotics of Eq. (\ref{eq:hI2expfun}) can be found by expanding $h(\alpha)$ to second order around its minimum, and substituting $\alpha=\alpha^*$ in the pre-exponential term.
One finds
	\begin{align}
	I_{\alpha\sim1}& \simeq \frac{bV}{2\sqrt{2\pi}} \left. \sqrt{\alpha\frac{1+2\alpha-\sqrt{1+4\alpha}}{1+4\alpha-\sqrt{1+4\alpha}}}\right\vert_{\alpha=2} e^{-\frac{1}{2} V^2 h(2)} \int_{\alpha_l}^{\alpha_h} e^{-\frac{1}{4} V^2 \left. \frac{\partial^2 h}{\partial \alpha^2}\right. \vert_{\alpha=2} (\alpha-2)^2} d\alpha \simeq \nonumber \\ &\simeq \frac{b V}{2\sqrt{3\pi}} \int_{-\infty}^{\infty}d\alpha e^{-\frac{1}{12} V^2(\alpha-2)^2} = b.
	\end{align}
In the transition from the first to the second line we extend the limits of integration to $\pm \infty$ because this action makes only asymptotically small corrections.
	
Crucially, this analysis demonstrates that the contribution from moments with $\alpha<\alpha_l$, and in particular with $\alpha>\alpha_h$, to the sum (\ref{eq:momseries}) is sub-dominant when $V \gg 1$. The only exception is the zero'th moment, whose value is always $1$. By adding the two contributions, we find
\begin{equation}
I\simeq 1+ I_{\alpha\sim1}=1+b,
\end{equation}
which is the correct result. The reason for our choice of $b<2$ should be clear now. It ensures that the dominant contributions to the moment series come from moments whose asymptotic behavior is captures by Equation (\ref{eq:momentapm=v}). Considering also processes with $b>2$ would have required a more refined asymptotic analysis of higher moments. Such analysis is not needed for the purpose of showing an example of non-uniform convergence, as the higher moments' behavior would be identical.

The considerations above show how non-uniform convergence is expressed in the series of moments. Each individual moment satisfies $\lim_{V \rightarrow \infty} M_n =0$.
However, moments of order that scales as $n \simeq V^2$ can be large enough to contribute to the series (\ref{eq:momseries}). The appearance of the factor $1/n!$ is important in determining the dominant part of the sum, as well as in suppressing the contribution from even higher orders. For $b<2$, the moments with $\alpha \simeq 2$ are these that restore the Jarzynski equality. The calculation above shows that the dominant moments are of orders that lie in a region of width $\delta n \sim \sqrt{6} V$ around $n*=2 V^2$. As $V$ is increased this dominant region is pushed to higher and higher moments and becomes wider. This behavior is the cause for the inability to exchange the order between taking the limit and the summation over moments. This is the same qualitative behavior that was found for the information engine in Section \ref{sect:vanishingerror}.
	
\label{sect:idealgas}
\section{Conclusions}

In this work we highlight an interesting property of the moment expansion of exponential averages, namely the occurrence of non-uniform convergence in certain physically interesting limits. The non-uniform behavior of the moments appears in situations where the limit of interest does not commute with other operations, such as summation over moments or integration over the thermodynamic observable of interest. We give two quite different examples for this behavior. We first consider a model of a simple information engine, which exhibits non-uniform convergence in the limit of error-free measurements. We then study a model of a gas particle in an expanding piston. Here the interesting limit is that of an infinitely large piston velocity. While the importance of order of limits have been pointed out before \cite{presse2006ordering,gu}, our results highlight a different aspect associated with such singular limits.

The non-uniformity manifests itself through the asymptotic behavior of the series of moments. For each value of $y \simeq y_0$ there is a range of moments whose distance from their limiting value is small. At the same time, there is a range of moments with $\left| M_n (y) - M_n (y_0)\right| \gg 1$. As $y$ approaches $y_0$ the transition between the two regions is pushed towards higher and higher moment orders, as seen in Figure \ref{fig:logmomentsVSnVSerror}. Naively taking the $y \rightarrow y_0$ limit, and replacing each moment by its limiting value $M_n (y_0)$, may result in an apparent violation of the associated integral work relation. 

Taking a closer look at the moment expansion of the exponential average, one finds that the sum is dominated by a well-defined group of moments. This group is responsible to the validity of the integral work relation when the limit of interest is taken. The factor of $1/n!$ has a crucial role in determining these dominant moments; it helps to create a peak that is centered around moments of order $n^* (y)$. In both examples we studied, the center of this region, $n^* (y)$, was pushed towards arbitrarily large values as $y \rightarrow y_0$. Interestingly, in both cases the width of this dominant region scales as $\sqrt{n^*}$. This is consistent with moment whose asymptotic behavior is dominated by an expression of the form $M_n \sim f^n(y)$. The non-uniform behavior of the moments helps to illuminate the importance of order of mathematical operations. Naively taking the $y \rightarrow y_0$ limit first assigns each moment its limiting value, which ultimately misses the contributions due to deviations from this limiting behavior. Crucially, the latter includes dominant moments necessary for convergence to the correct value of the exponential average.

Although we demonstrated the non-uniformity on just two models, we believe this property is more general and can easily be found in other processes. It will be interesting to search for similar phenomena when considering limits in which particle masses, or observation times, approach infinity. 
	\label{sect:discussion}
	

%

\end{document}